\documentclass[aps,prl,showkeys,tightenlines,nofootinbib, twocolumn]{revtex4-1}
\usepackage{comment}

\usepackage{amsmath,amssymb, amsfonts}
\usepackage{graphicx}

\begin{document}

\title{Graviton and gluon scattering from first principles} 
 
\author{Rutger H.  Boels}
\email{Rutger.Boels@desy.de}
\affiliation{II. Institut f\"ur Theoretische Physik, Universit\"at Hamburg,  Luruper Chaussee 149, D-22761 Hamburg, Germany}
\author{Ricardo Medina}
\email{ rmedina@unifei.edu.br}
\affiliation{Instituto de F\'\i sica,
                 Universidade de S\~ao Paulo,\\
                 Rua do Mat\~ao 1371, 05508-090 S\~ao Paulo - SP,  Brazil}
\altaffiliation{on sabbatical leave}
\affiliation{	Instituto de Matem\'atica e Computa\c{c}\~ao, 
                 Universidade Federal de Itajub\'a,
                 Av. BPS 1303, 
                 37500-903, Itajub\'a - MG, Brazil}
\altaffiliation{permanent address}


\begin{abstract}
\noindent
Graviton and gluon scattering are studied from minimal physical assumptions such as Poincare and gauge symmetry as well as unitarity. The assumptions lead to an interesting and surprisingly restrictive set of linear equations. This shows gluon and graviton scattering to be related in many field and string theories, explaining and extending several known results. By systematic analysis exceptional graviton scattering amplitudes are derived which in general dimensions can not be related to gluon amplitudes. The simplicity of the formalism guarantees wide further applicability to gauge and gravity theories. 
\end{abstract}





\maketitle

\section{Introduction}
Scattering experiments are the bedrock of physical understanding at the smallest scales. A key ingredient in predictions for these are scattering amplitudes, which are a measure for the likelihood of a given scattering process. A key step in the chain from theory to experiment is therefore the computation of these amplitudes, and hence they are a central object of study in almost any proposed theory of nature. 

General relativity effects can usually be ignored on collider-accessible scales due to the extraordinary weakness of gravity compared to the other known forces of nature encapsulated in the standard model of particle physics. However, on very small scales in cataclysmic events such as colliding black holes \cite{Abbott:2016blz} or big bang cosmology gravity becomes important. Physics such as that of black holes force one to consider quantum theories of gravity \cite{Woodard:2009ns}. These motivations drive the study of scattering amplitudes of the spin-$2$ force-particle of gravity: the graviton. These describe colliding gravity waves in a flat background driven by the non-linear nature of general relativity itself.

Graviton scattering amplitudes are notoriously complicated to compute even without quantum corrections, see e.g. \cite{'tHooft:1974bx}. In practice these amplitudes are computed using known perturbative relations to the much simpler scattering amplitudes of spin-$1$ particles dubbed gluons. Gluons are the force-particle present in a class of models known as (Yang-Mills) gauge theory. The standard model of particle physics is for instance in this class. The first relations between gluon and graviton scattering were derived in string theory \cite{Kawai:1985xq}, whose particle-limit yields relations between Einstein-Hilbert gravity and pure gauge theory. The physical origin of these relations has never been clarified. Closest is a field theory-based proof for Einstein-Hilbert gravity \cite{BjerrumBohr:2010ta}. 

New light was shed on this gauge-gravity connection in field theory in  \cite{Bern:2008qj,Bern:2010ue} that introduced the notion of color-kinematic duality, assigning a group theory structure to the momentum dependent parts of amplitudes. This induces relations between gluon amplitudes (proven in \cite{BjerrumBohr:2009rd, Stieberger:2009hq}) as well as relations between gauge and gravity amplitudes, see \cite{Bern:2010yg}. Color-kinematic duality has been instrumental in pushing the computation of quantum corrections in gravity theories, see e.g. \cite{Bern:2009kd}, revealing surprising cancellations for ultraviolet divergences. Extensions to other solutions of the field equations have been explored \cite{Monteiro:2014cda} as well as first steps toward effective field theory \cite{Broedel:2012rc}. A systematic understanding of the physical origin of color-kinematic duality is however absent.

In this article we provide a first-principles based approach to gluon and graviton scattering and relations between these processes. These relations are shown to be a direct consequence of linearised diffeomorphism symmetry in examples. This mirrors \cite{Barreiro:2013dpa} where gluon amplitude relations were shown to result from linearised gauge symmetry. Our analysis also clearly shows some of the limitations: there are gravitational amplitudes which cannot be related to sums over products of gluon amplitudes already at four points. 

\section{Physical constraints}
In this letter $D$-dimensional, parity even, flat space scattering amplitudes are studied of up to spin two massless particles with integer spins and complex momenta. That is, for each external particle (``leg'') of the scattering amplitude the data is
\begin{equation}
p_i^{\mu} \qquad p_i^2 = 0 \qquad \textrm{Irrep } SO(D-2) 
\end{equation}
Translation invariance leads to momentum conservation,
\begin{equation}
\sum_i p^{\mu}_i = 0
\end{equation}
The irreducible representation (irrep) of the little group $SO(D-2)$ is specified by a product of polarisation vectors $\xi^{I}_{\mu}$ where $I$ is the little group and $\mu$ the Lorentz group index (see e.g. \cite{Boels:2009bv}). For spin-one gluons, a single polarisation vector is needed, while the tensor product of two polarisation vectors can be decomposed into anti-symmetric, traceless-symmetric and trace irreps. Physically, these correspond to the 2-form, the graviton and the dilaton respectively. The scattering amplitude must be multi-linear in the polarisation vectors, i.e.
\begin{equation}\label{eq:multilin}
A = \xi_{1,\mu_1} \ldots \xi_{n-1,\mu_{n-1}} \xi_{n,\mu_n} \left(I \right)^{\mu_1 \ldots \mu_{n-1} \mu_n}
\end{equation}
Since the scattering amplitude is a scalar and parity symmetry is assumed, the tensor $I$ must be built from momenta and metrics. Amplitudes are assumed to have a homogeneous mass dimension without loss of generality. 

To reduce the physical degrees of freedom from $D$ to the physical value of $D-2$, polarisation vectors must obey two constraints. First, they are transverse:
\begin{equation}
p_i \cdot \xi_i = 0
\end{equation}
Second, replacing any polarisation vector by the momentum in the leg leads to a vanishing amplitude:
\begin{equation}
A(\ldots \xi_i \rightarrow p_i \ldots ) = 0
\end{equation}
This is called on-shell gauge invariance. It can be understood as a consequence of the Schwinger-Dyson equation for amputated, on-shell legs, or equivalently as a consequence of Noethers' second theorem \cite{Noether:1918zz}. Both arguments restrict to standard, quadratic propagators for unitarity and causality. Perturbative unitarity furthermore restricts for instance kinematic poles and their residues at tree level. 

Finally, the scattering amplitudes must be symmetric under interchange of the quantum numbers of any two indistinguishable, integer spin particles. The quantum numbers include momenta, polarisations and any internal symmetry indices: this is the most general possibility \cite{Coleman:1967ad}. Gravitons have no additional internal symmetry indices, but gluons can have an internal, adjoint-valued Lie group index. Matter can be in general representations of an internal symmetry. In this letter the focus is mostly on the kinematics. The listed constraints are very general and apply to string and field theories alike. 

\section{Solving the constraints: general}
An important building Lorentz-invariant block are contractions of two momenta, $p_i \cdot p_j$. By momentum conservation these are an overcomplete set, and a (cyclic) basis for $n$ particles can be chosen as
\begin{align}
(p_i + p_{i+1})^2 & \qquad i = 1, \ldots, n\\
(p_i + p_{i+1}+p_{i+2})^2 &  \qquad i = 1, \ldots,n\\
\ldots\\
(p_i + \ldots+p_{i+[n/2]})^2  & \qquad i= 1, \ldots \mathcal{N}
\end{align}
where $[n/2]$ is the nearest integer smaller or equal to $n/2$, $\mathcal{N}$ is $n/2$ or $n$ for $n$ even or odd and momenta are identified cyclicly. All inner products of momenta can be expressed in this set of $\frac{1}{2} D (D-3)$ elements. The elements of the basis set will be referred to as ``Mandelstam invariants''. 

The tensor $I$ in equation \eqref{eq:multilin} is built from metrics and momenta. For the parity-odd case (not considered here), a single Levi-Cevita tensor would be added. First use momentum conservation to eliminate $p_n$ from any tensor $I$. Then, for $i\neq n$, 
\begin{equation}
p_j \cdot \xi_i \qquad j \neq i, n 
\end{equation}
are all possible contractions. For $i=n$, one can choose
\begin{equation}
p_j \cdot \xi_n \qquad j \neq 1, n 
\end{equation}
to implement transversality. Any contraction of external momentum with polarisation vectors can be expressed in this set. It is useful to classify by the number of metric contractions between polarisation vectors.

Since the scattering amplitudes are multi-linear in polarisation vectors, any amplitude is a linear combination of all possible metric and momentum contractions with the polarisation vectors after solving momentum conservation and transversality. In other words, amplitudes in this language can be treated as vectors in a vector space. Amplitude relations with helicity-blind coefficients appear as vector relations in this space. This is a powerful observation, made in  \cite{Barreiro:2013dpa} for gluon amplitudes.

A remaining constraint is on-shell gauge invariance for each external leg separately. The polarisation tensor building blocks after this constraint for a given leg can be related to a minimal set again, resulting in a high number of \emph{linear} equations for the coefficients of the general amplitude vector. Repeating for all external polarisation vector gives a very large, dependent set of linear equations. The dimension of the null-space of the resulting matrix is the number of independent solutions to the on-shell constraints. Every solution to the on-shell constraints can be expressed in a basis which contains this number of independent elements, with coefficients that are functions of Mandelstam invariants.  

\section{Solving the constraints: results}
As a first example, consider the three point amplitude for three gluons. For three particles, there are no Mandelstam invariants. The $4$ building blocks are:
\begin{multline}
\vec{I} = \left\{ (\xi_1 \cdot \xi_2 )(p_2 \cdot \xi_3), (\xi_1 \cdot \xi_3 )(p_1 \cdot \xi_2), \right. \\ \left. (\xi_2 \cdot \xi_3 )(p_2 \cdot \xi_1),(p_2 \cdot \xi_1)(p_1 \cdot \xi_2)(p_2 \cdot \xi_3)  \right\}
\end{multline}
three of which contain a single metric contraction. A scattering amplitude is a linear combination of these blocks, $A = \vec{\alpha} \cdot \vec{I}$. Setting $\xi_3 \rightarrow p_3$ gives
\begin{multline}\nonumber
\vec{I}\lfloor_{\xi_3 \rightarrow p_3} = \left\{ 0, - (p_2 \cdot \xi_1  )(p_1 \cdot \xi_2), - (p_1 \cdot \xi_2  )(p_2 \cdot \xi_1),0 \right\}
\end{multline}
where momentum conservation was used to re-write in terms of the chosen basis. Repeating for the other two legs gives a matrix equation,
\begin{equation}\nonumber
\left(\begin{array}{cccc} 0 & -1& - 1&0 \\ 1 & 0 & 1&0\\ 1 & -1& 0&0 \end{array}\right)  \vec{\alpha} = 0
\end{equation}
which is a rank $2$ matrix. Hence, there are two possible gluon scattering amplitudes proportional to either
\begin{equation}\label{eq:solsYM}
\vec{\alpha} = \{-1,-1,1,0\} \quad \textrm{or} \quad \vec{\alpha} = \{0,0,0,1\}
\end{equation}
These are readily identified as the Yang-Mills amplitude and the amplitude generated by an $F^3$-type interaction respectively, up to the coupling constant. For three gravitons, it is easier to first solve the on-shell constraints for generic products of two polarisation vectors and specialise to gravitons later. New here is the appearance of scalar factors, $\xi^I_i \cdot \xi^J_i \propto \delta^{IJ}$
which will be set to zero in the remainder to avoid double counting. Solving the on-shell constraints for three gravitons then yields four independent solutions. A special class of solutions to the on-shell constraints can be obtained by multiplying solutions for gluon amplitudes, for which there are also four solutions
\begin{equation}\label{eq:threepointsquares}
\left\{A^L_{\textrm{YM}} A^R_{\textrm{YM}}\,,\, A^L_{F^3} A^R_{\textrm{YM}}\,,\, A^L_{\textrm{YM}} A^R_{F^3}\,,\,A^L_{F^3} A^R_{F^3}\right\}
\end{equation}
which can be checked to give independent vectors. Graviton amplitudes arise from symmetrising over `left' and `right' polarisation vectors, giving three possibilities. Hence, products of gluon amplitudes furnish a basis for all graviton three point amplitudes. The first element in the set of products is proportional to the Einstein-Hilbert answer. Note that the gravity solutions have a minimal number of two, one or no metric contractions. 

By unitarity, theories with these three point amplitudes must have higher point amplitudes, enumerated by coupling constants. Consider first the field theory built on the first element in equation \eqref{eq:threepointsquares}, i.e. with a single coupling constant. By dimensional analysis and unitarity the maximal amount of momenta in the numerator of an $n$ point graviton amplitude in this theory is $2 n- 4$. This implies there are here always at least two metric contractions between polarisation vectors for this solution to the on-shell constraints consistent with unitarity. A similar argument applies to Yang-Mills theory at tree level. For maximal supersymmetric theories such as string theory these metric-counting criteria certainly hold for four and five points and likely hold in general already from first principles alone, see \cite{Elvang:2009wd}. For non-maximal \cite{Mafra:2016nwr, Berg:2016wux} or no \cite{Bern:1993qk} supersymmetry counterexamples are known at loop level. 

By solving the on-shell constraints, the tables \ref{table:gluons} and \ref{table:gravitons} were obtained. For graviton legs the symmetric representation was singled out, but traces not explicitly subtracted. For both gravitons and gluons Bose symmetry will be analysed separately below. The explicit computation was done in Mathematica, with help from the {\tt linbox} library beyond the double metric terms at five points for gravitons and beyond six points for gluons. In these cases only the dimension of the solution space was computed using integer numerics. Note the computational complexity increases rapidly with number of particles, as well as the striking difference between the numbers of solutions with or without constraints on the number of metrics. 

\begin{table}[ht]
\caption{solving on-shell constraints: gluons} 
\centering
\begin{tabular}{c | c c c} 
\hline 
\# gluons &  min \#   & \# terms in &  \# on-shell gauge \\
& metrics & Ansatz & invariant\\
\hline 
4  & 1 & 27 & 1 \\ 
    & 0 & 43 & 10 \\ 
5  & 1 & 315 & 2 \\
    & 0 & 558 & 142  \\
6  & 1 & 4575 & 6 \\
    & 0  & 8671 & 2364  \\
7  & 1 & 79275 & 24 \\
    & 0  & 157400 & 45028  \\
8  & 1 & 1593753 & 120
\end{tabular}
\label{table:gluons}
\end{table} 

\begin{table}[ht]
\caption{solving on-shell constraints: gravitons} 
\centering
\begin{tabular}{c | c c c} 
\hline 
\# gravitons &  min \#   & \# terms in &   \# on-shell gauge \\
& metrics & Ansatz & invariant\\
\hline 
4 & 2 & 5046 & 1 \\ 
   & 1 & 6838 & 11 \\ 
   & 0 & 7094 & 56 \\ 
5 & 2 & 27921 & 3 \\
   & 1 & 47362 & 692 \\
   & 0 & 55138 & 5555  
\end{tabular}
\label{table:gravitons}
\end{table} 

Since Yang-Mills theory and open superstring theory yield tree-level gluon scattering amplitudes with at least a single metric contraction, the numbers in the tables imply  up to eight points that if $(n-3)!$ independent amplitudes form a basis in these theories. Any amplitude may be expressed in this set by relations with coefficients that are functions of Mandelstam invariants  \cite{Barreiro:2013dpa}. For color-ordered amplitudes in Yang-Mills theory these are the BCJ relations, while for string theory these are the monodromy relations \cite{Plahte:1970wy, BjerrumBohr:2009rd, Stieberger:2009hq}. The form of the BCJ relation(s) can be derived from matrix algebra where explicit solutions are available.

Physical scattering amplitudes are specific linear combinations with Mandelstam-dependent coefficients of the explicit solutions of the constraints. For four gluons for instance, the single single-metric-contraction result can be transformed to a unique local and Bose-symmetric form, 
\begin{equation}
 I\left(\xi_1, \xi_2, \xi_3, \xi_4\right)
\end{equation}
where internal symmetry indices were disregarded. This has minimally mass dimension four, while the Yang-Mills amplitude consistent with factorisation for the first solution in equation \eqref{eq:solsYM} has mass dimension zero. Double poles are not allowed, so the only possibility consistent with a cyclic color-ordered amplitude $A(1,2,3,4)$ is
\begin{equation}\label{eq:colordYM4pt}
A^{\textrm{co}}_{YM}(1,2,3,4) = \frac{ I\left(\xi_1, \xi_2, \xi_3, \xi_4\right)}{(p_1 + p_2)^2 \, (p_2+p_3)^2} 
\end{equation}
up to a proportionality constant which is fixed from the three point amplitude by unitarity. A similar computation yields explicit color-ordered Yang-Mills amplitudes up to six points from first principles, see also \cite{Schuster:2008nh}. 

On the gravity side, the above table shows the existence of relations between graviton and gluon amplitudes at four and five points, and some limitations. Special solutions to the gravitational on-shell constraints arise from all possible products of gluon amplitudes, with the result symmetrised over both gluon polarisations associated to each graviton leg. As a special case, the result is symmetric under complete swaps of both `left' and `right' gluons. Since Einstein-Hilbert gravity and closed superstring theory yield tree-level graviton scattering amplitudes with at least two metric contractions, products of gluon amplitudes furnish a \emph{complete} basis for the space of solutions to the on-shell constraints for gravitons, to the level checked. This proves the existence of relations with Mandelstam dependent coefficients between graviton amplitudes and products of gluon amplitudes up to and including five points. The exact relations that result at tree level are easily identified as the (field theory limit of the) appropriate KLT relations. 

For four gravitons, the natural starting point is the left-right symmetrised product of gluon solutions,
\begin{equation}
\left[ I\left(\xi^L_1, \xi^L_2, \xi^L_3, \xi^L_4\right)  I\left(\xi^R_1, \xi^R_2, \xi^R_3, \xi^R_4\right) \right]_{L \leftrightarrow R}
\end{equation}
The gravity amplitude is proportional to this factor by Table \ref{table:gravitons}. The unique possibility consistent with symmetry, unitarity and the three point amplitude is
\begin{equation}
A_{EH} (1,2,3,4)  =  \frac{ \left[  I^L I^R\right]_{L \leftrightarrow R}}{(p_1 + p_2)^2 \, (p_2+p_3)^2 \, (p_1 + p_3)^2}
\end{equation}
The five point Einstein-Hilbert amplitude was computed similarly by imposing unitarity, uniform mass dimension and Bose symmetry. Solving the on-shell constraints for six graviton amplitudes proved beyond our current setup. These are the first computations of non-trivial graviton scattering in Einstein-Hilbert theory from first principles alone.

In general, the conjectured number of independent solutions to the on-shell constraints for gluons with a single metric contraction is $(n-3)!$, while this number for gravitons with a double metric contraction is $\leq ((n-3)!) ((n-3)! +1) / 2$. Symmetrised products of gluon amplitudes are expected to form a basis for the solutions to the on-shell constraints for gravity: this will yield gauge-gravity relations of the KLT-type for all multiplicity. The techniques displayed here can easily be extended to explore more general theories involving generalised matter content. The simplest class are amplitudes involving two-form and dilatons on the gravitational side. Only slightly more complicated are amplitudes with scalars on the gluon side (see e.g. \cite{Gehrmann:2011aa}): these can be related on the gravitational side to amplitudes with vector or scalar matter. Moreover, the existence of relations between pure gluon and mixed gluon/graviton amplitudes (see e.g.  \cite{Stieberger:2014cea}) arise from solving the same set of on-shell constraints.

\section{Gravitational amplitudes that cannot be related to gluons}
The counting in the table above indicates that there are graviton amplitudes that cannot be written as sums over products of gluon amplitudes.  At four points with one or more metric contractions for instance there are eleven independent solutions, but only ten of these may be obtained as a product of the unique Yang-Mills amplitude (e.g. \eqref{eq:colordYM4pt}) as one of the factors. For a more precise analysis, Bose-symmetry must be studied first. This reveals an interesting substructure in the gluon terms, which is of independent interest. 

Some experimentation with explicit symmetrisation gave seven independent, local gluon amplitudes which are completely symmetric. Their mass dimensions are 
\begin{equation}
\textrm{min} [mass] = 4, 4, 6, 6, 6, 8, 8  \qquad \textrm{symmetric}
\end{equation}
where overall powers of the two basis completely symmetric polynomials, $\sigma_1 = s^2+t^2+u^2$ and $\sigma_2 = s\,t\,u$, have been factored out. The remaining three solutions can be taken to be anti-symmetric in three chosen legs. Their minimal mass dimensions are
\begin{equation}
\textrm{min} [mass] = 8, 10, 12 \qquad \textrm{non-symmetric}
\end{equation}
This class of amplitudes cannot be symmetrised since any polynomial of Mandelstams anti-symmetric under exchange of three legs is proportional to the unique completely anti-symmetric polynomial $\propto (u-s)(s-t)(t-u)$ by considering its roots. The natural gauge group for these amplitudes contain a $U(1)$ factor: these amplitudes likely never occur in gauge theories with adjoint matter only. From the symmetry properties of gluon amplitudes it is easy to obtain a basis for those graviton scattering amplitudes that can be related to sums over products of gluon amplitudes: 
\begin{multline}
\left[(10) \otimes (10)\right]_{\textrm{S}} = \left[(1) \otimes (1)\right]_{\textrm{S}} + \left[(1) \otimes (7)\right]_{\textrm{S}}  \\ + \left[(7) \otimes (7)\right]_{\textrm{S}} + \left[(3) \otimes (3)\right]_{\textrm{S}} 
\end{multline}
where the subscript stand for both Bose-symmetry as well as left-right symmetry. We have checked that after the symmetrisations there are $1$, $7$ and $31$ independent terms for up to two, one and no metric contractions amplitudes respectively.

In the single metric case $8$ independent solutions are obtained from Bose-symmetrisation of the gravitational solution to the on-shell constraints, projecting out the same three solutions as in the gluon case. This leaves in this class a single, Bose-symmetric amplitude which cannot be written as a sum over products of gluon amplitudes. Note that this exceptional four dimensional amplitude has vanishing helicity-equal and single-helicity-un-equal amplitudes, a prerequisite for supersymmetry \cite{Grisaru:1977px}. In the more general case four non-squaring $4$ point amplitudes are obtained after Bose-symmetrization, of minimal mass dimension 
\begin{equation}
\textrm{min} [mass] = 12,16,18,18
\end{equation}
The first entry is the `at least single metric' amplitude. In addition an amplitude with three gravitons and a scalar was found which cannot be written as a sum over products of single scalar and $3$ gluon amplitudes. Table \ref{table:gravitons} indicates many more examples at the five point level. 

It should be stressed that in specific dimensions exceptions to the analysis above may arise. For instance, multiplying two parity odd gluon amplitudes yield a parity even gravity amplitude. Furthermore, subtracting the traces in general is dimension dependent, but may lead to special cancellations in specific dimensions. Finally, in fixed dimensions special kinematic (Gramm determinant) relations hold. For instance, only five out of seven symmetric four gluon factors are independent in strictly four dimensions and, as expected by electric-magnetic duality, only one in three.

\subsection{Discussion}
In this letter a first principles approach to gravity and gluon scattering was introduced. This provides a physical understanding of the origin of several known relations between these a-priori seemingly different processes: they are deeply rooted in the gauge invariances of the gluon and graviton. As shown explicitly not all gravitational theories obey such relations however. Note that the amplitudes in these exceptional theories cannot arise in closed string theory at tree level; their absence is a string theory prediction. They could arise as counterterms in string or field theory, where they pose a particular challenge to color-kinematic duality.  The simplicity of the assumptions guarantees wide applicability. 

Extending the explicit analysis to general multiplicities as well as matter content is a priority, including obtaining explicit amplitudes. Dimensional constraints should be studied, also in connection with fermionic matter in various representations of the gauge group. Tracing our results to the effective action has direct connections to the effective action approach for beyond-the-standard-model physics, see e.g. \cite{Cheung:2015aba} and references therein, as well as to gravity, see e.g. \cite{BjerrumBohr:2003af} and string theory, see e.g.  \cite{Barreiro:2012aw}. Curved backgrounds are a further target to be studied. 

The techniques employed here have immediate practical uses beyond the tree level  \cite{Gehrmann:2011aa} \cite{Glover:2003cm}. At minimum, they provide the minimal dimension of the basis of helicity dependent parity-even factors to all loop orders, but with coefficients which are generically non-rational functions of the momenta. These are an interesting target. Also, (generalised) unitarity driven methods combine directly with the techniques presented here: already at one loop this has the potential to provide significant saving. Further research directions include exploring physical consequences of generic constraints, see e.g. \cite{Benincasa:2007xk}. Also, the string KLT relations extend to the full spectrum of the open and closed string; the analog of on-shell gauge invariance which yields these more general relations should be highly interesting.

\section*{Acknowledgements}
The authors would like to thank Luiz Antonio Barreiro, Hui Luo, Sven-Olaf Moch and Oliver Schlotterer for discussions. RM is grateful to Hamburg University and DESY for their hospitality.  RB would like to thank the Isaac Newton Institute for Mathematical Sciences, Cambridge, for support and hospitality during the programme Gravity, Twistors and Amplitudes. This work was supported by EPSRC grant no EP/K032208/1 and by the German Science Foundation (DFG) within the Collaborative Research Center 676 ``Particles, Strings and the Early Universe".

\bibliographystyle{apsrev4-1}

\bibliography{squaring}

\end{document}